%% file: icc.tex
\documentclass[conference,10pt]{IEEEtran}
 \IEEEoverridecommandlockouts
\usepackage{cite}
\usepackage{color}
\usepackage[pdftex]{graphicx}
\graphicspath{{fig/}{jpeg/}}
\usepackage[cmex10]{amsmath}
\usepackage{amssymb, bbm}
\usepackage{algorithm}
\usepackage{epstopdf}
\usepackage{algorithmic}
\usepackage{subcaption,environ}
\usepackage{hyperref}
\usepackage{comment}
\input{mysymbol.sty}
\input{my_sections.tex}

\usepackage{tikz}
\usetikzlibrary{shapes,arrows}
\usetikzlibrary{matrix} % for block alignment
\usetikzlibrary{decorations.markings} % for arrow heads
\usetikzlibrary{calc, patterns} % for manimulation of coordinates
\usetikzlibrary{decorations.pathmorphing,snakes, positioning, decorations.pathreplacing}

\usetikzlibrary{fit}

\makeatletter
\newsavebox{\measure@tikzpicture}
\NewEnviron{scaletikzpicturetowidth}[1]{%
  \def\tikz@width{#1}%
  \begin{lrbox}{\measure@tikzpicture}%
  \BODY
  \end{lrbox}%
  \pgfmathparse{#1/\wd\measure@tikzpicture}%
  \BODY
}
\makeatother

\usepackage{pgfplots}
\usepackage{color}
\usepackage{multirow}

\newcommand{\QED}{\hfill\ensuremath{\blacksquare}}

\def\E{\mathbb{E}}

\def\Tr{\text{Tr}}

\begin{document} 

\title{Scheduling Low Latency Traffic for\\ Wireless Control Systems in 5G Networks}
\author{Mark Eisen$^{\dagger}$ \quad Mohammad M. Rashid$^{\dagger}$ \quad Alejandro Ribeiro$^{*}$ \quad
			Dave Cavalcanti$^{\dagger}$
\thanks{{Supported by Intel Science and Technology Center for Wireless Autonomous Systems. The authors are with $^{\dagger}$Intel Corporation, Hillsboro, OR 97124 and $^{*}$University of Pennsylvania, Philadelphia, PA 19104. Emails: \{mark.eisen, mamun.rashid, dave.cavalcanti\}@intel.com} and \{aribeiro\}@seas.upenn.edu.}}
\maketitle

\begin{abstract}
We consider the problem of allocating 5G radio resources over wireless communication links to control a series of independent low-latency wireless control systems common in industrial settings. Each control system sends state information to the base station to compute control signals under tight latency requirements. Such latency requirements can be met by restricting the uplink traffic to a single subframe in each 5G frame, thus ensuring a millisecond latency bound while leaving the remaining subframes available for scheduling overhead and coexisting broadband traffic. A linear assignment problem can be formulated to minimize the expected number of packet drops, but this alone is not sufficient to achieve good performance. We propose an optimal scheduling with respect to a control operation cost that allocates resources based on current control system needs. The resulting control-aware scheduling method is tested in simulation experiments that show drastically improved performance in 5G settings relative to control-agnostic scheduling under the proposed time-sliced frame structure.
\end{abstract}


\section{Introduction}
The increasing scale of modern IoT and industrial control systems has motivated the development of wireless control system technology that can achieve reliable performance in these settings~\cite{varghese2014wireless, li2017review,wollschlaeger2017future}. One of the primary challenge of wireless control in industrial settings, however, is the time sensitive nature of the systems, thus requiring low latency wireless transmissions~\cite{varghese2014wireless}.  The noise of the wireless channel makes it difficult to simultaneously maintain high reliability while achieving low latency. This motivates the design of resource allocation and scheduling strategies that can both meet reliability \emph{and} latency requirements of the industrial control system.

For such time-sensitive applications, generic delay-aware schedulers have been developed, such as EDF~\cite{wu2014analysis}, WFQ~\cite{lu1999fair}, and M-LWDF \cite{andrews2001providing}. However, recent developments and advancements in 5G cellular service \cite{parkvall2017nr} has created opportunity to meet latency and reliability standards for IoT systems in so-called ultra reliable low latency communications (URLLC) \cite{singh2016ultra, mahmood2016radio, li20175g, popovski20185g, anand2018joint}. Such works look specifically at the flexible frame structure defined by the 5G standard to develop URLLC techniques that may be employed in modern cellular environments. None of these works, however, consider the underlying control systems that compose these IoT systems in making decisions for scheduling or radio resource allocation.

Specifically in the context of wireless control systems, dynamic schedulers use control system information to provide access to the communication medium dynamically at each step. %, for example by dynamically assigning priorities to the competing tasks.
Resulting ``control-aware'' scheduling approaches make decisions explicitly based on current control system states ~\cite{ Cervin_event_scheduling, mamduhi2014event,shi2011optimal,han2017optimal} or wireless channel conditions~\cite{GatsisEtal15,ma2019optimal}. These works, however, do not consider the low-latency requirements of industrial or IoT systems. A control-aware scheduling design for low-latency systems has previously been developed for next-generation WiFi \cite{eisen2019control} but has not yet been considered in cellular systems. 

In this paper, we propose a scheduling framework for low-latency control system traffic in 5G communication systems that specifically adapts to the underlying control system needs. We model the switched system dynamics of a wireless control system and use this to model a control-optimal scheduling problem. Using just a simple separation of low-latency traffic for wireless control systems and standard broadband coexisting in the network---i.e. ``time'' slicing---we demonstrate how control-awareness allows us to effectively schedule low-latency traffic in just a single subframe. Numerical results demonstrate the benefit of control-optimal scheduling over standard throughput-based scheduling design.

\section{Problem Formulation}\label{sec_problem}

%%%%%%%%%%%%%%%%% F I G U R E %%%%%%%%%%%%%%%%5
%%%%%%%%%%%%%%%%%%%%%%%%%%%%%%%%%%%%%%%
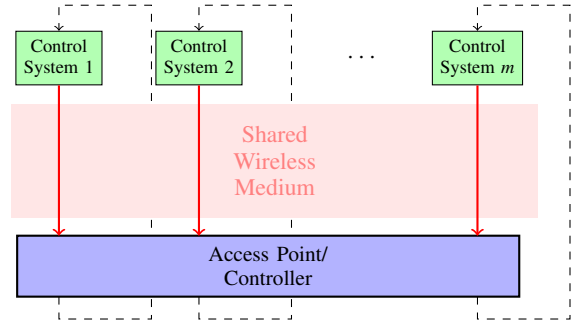
\begin{figure}
\centering
\input{tikz_wncs4}
\caption{Wireless control system with $m$ independent systems. Each system contains a sensor that measure state information, which is transmitted to the controller over a wireless channel. The state information is used by the controller to determine control policies for each of the systems.The communication is assumed to be wireless in the uplink and ideal in the downlink.}
\label{fig_wcs}
\end{figure}

Consider a system of $m$ independent control systems as show in Figure \ref{fig_wcs}, where each system $i=1,\hdots,m$ maintains a state variable $\bbx_{i} \in \reals^p$. The dynamics evolve over a discrete time index $k$.  Given a control input $\bbu_{i,k} \in \reals^q$ , the state evolve based on the generic linear state space representation,
\begin{align}\label{eq_control_orig}
\bbx_{i,k+1} &= \bbA_{i} \bbx_{i,k} + \bbB_{i} \bbu_{i,k} + \bbw_k
\end{align}
where $\bbA_{i} \in \reals^{p \times p}$ and $\bbB_{i} \in \reals^{p \times q}$ are matrices that define the system dynamics, and $\bbw_{k} \in \reals^{p}$ is i.i.d. random noise with zero mean and co-variance $\bbW_i$ that captures the noise in the model. The plant dynamics in \eqref{eq_control_orig} is itself generic and can represent different systems found in IoT environments. Systems that have nonlinear dynamics can be modeled as approximately linear close to equilibrium points, with linearization error captured by $\bbw_k$. 

In modern IoT settings, such plants represent a set of devices, e.g. machinery or robotic systems, whose state is measured locally on-device but whose control signal input $\bbu_{i,k}$ is computed by a centralized processor/coordinator. This so-called control loop is thus closed through wireless communication between the plants and the centralized controller co-located at a common wireless 5G base station (BS). The dynamics are modeled here as discrete, but in reality the plants evolve continuously over time.  With fast dynamics and high sampling rates, many IoT devices require strict low latency between the time state information is measured by the plant to when it can be processed by the controller at the BS. Observe in Figure \ref{fig_wcs} that this model restricts its attention to wireless connections in uplink of the control loop, while downlink is assumed to occur over an ideal channel.

Aside from latency requirements, wireless communications is inherently noisy and prone to packet drops. Thus, to model the closed-loop dynamics of the plants, we must consider the effect of lost state information. To study the details, consider a generic linear control $ \bbu_{i,k} = \bbK_i \bbx_{i,k}$ for some matrix $\bbK_i \in \reals^{q \times p}$ that drives the state $\bbx_{i,k}$ to a desired operating point. When exact state information is not known to the BS due to packet drops, it estimates the state of device $i$ at time $k$ as
\begin{align}\label{eq_state_est}
\hbx^{(l_{i,k})}_{i,k} :=(\bbA_i + \bbB_i\bbK_i)^{l_{i,k}} \bbx_{i,k-l_{i,k}},
\end{align}
where $l_{i,k} \geq 1$ is a counter of successive failed transmissions. Observe that in \eqref{eq_state_est} we assume the BS/controller has knowledge of $\bbA_i, \bbB_i$, and $\bbK_i$, but not the noise $\bbw_k$. 

At time $k$, if the state information is received, the controller applies the input $\bbu_{i,k} = \bbK_i \bbx_{i,k}$ using the exact state, and otherwise applies input $\bbu_{i,k} = \bbK_i \hbx_{i,k}$. At each cycle $k$ we consider a binary variable $\gamma_{i,k} \in \{0,1\}$ that indicates whether the uplink transmission was successful. We obtain then the following switched system dynamics for $\bbx_{i,k}$ as
\begin{align}\label{eq_control_switch}
\bbx_{i,k+1} = \begin{cases}
(\bbA_i + \bbB_i \bbK_i) \bbx_{i,k} + \bbw_k, \ &\gamma_{i,k} = 1, \\
\bbA_i \bbx_{i,k} + \bbB_i\bbK_i\hbx^{(l_{i,k})}_{i,k} + \bbw_k, \ &\gamma_{i,k} = 0.
\end{cases}
\end{align}
The previous transmission counter $l_{i,k}$ is updated at time $k$ as
\begin{align}
l_{i,k+1} = \begin{cases}
1, \ &\gamma_{i,k} = 1, \\
l_{i,k} + 1, \ &\gamma_{i,k} = 0.
\end{cases} \label{eq_time_switch}
\end{align}
Observe that the successive error between the true and estimated state can be written as
$\bbe_{i,k} := \bbx_{i,k} - \hbx^{(l_{i,k})}_{i,k} = \sum_{j=0}^{l_{i,k}-1}(\bbA_i+\bbB_i \bbK_i)^{j} \bbw_{i,k-j-1}$. It is evident that this error grows with the transmission counter $l_{i,k}$. We proceed now to describe the 5G scheduling architecture and a framework for scheduling low-latency traffic to control IoT devices.

%%%%%%%%%%%%%%%%%%%%%%%%%%%%%%%%%%%%%%%%%%%%5
%%%%%%%%%%%%%%%%%%%%%%%%%%%%%%%%%%%%%%%%%%%%5
%%%%%%%%%%%%%%%%%%%%%%%%%%%%%%%%%%%%%%%%%%%%%%
\subsection{5G scheduling architecture}\label{sec_comm_model}

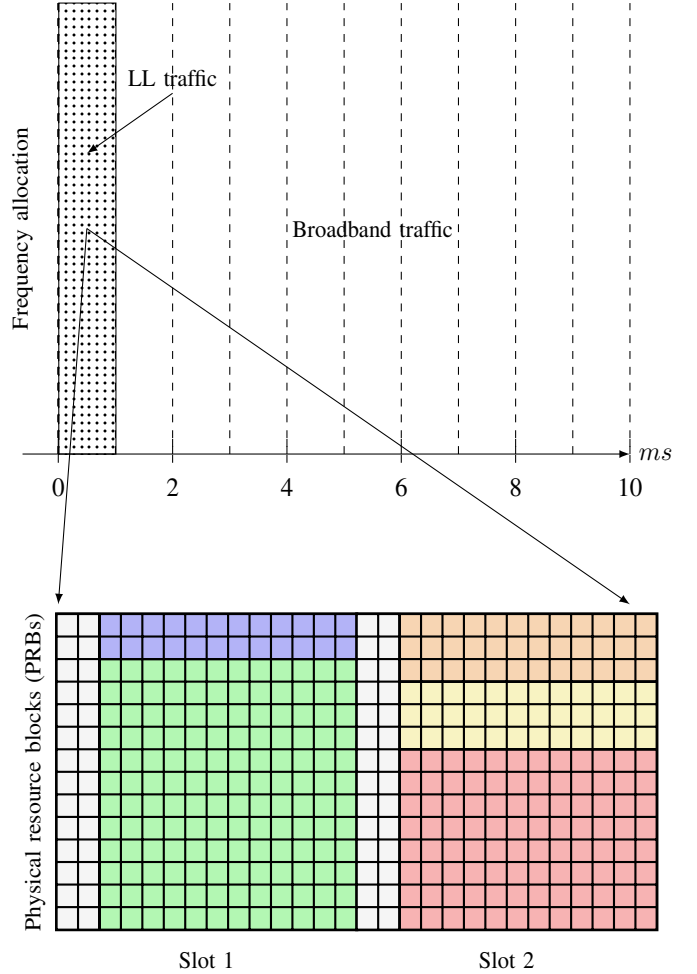
\begin{figure}
\centering
\begin{scaletikzpicturetowidth}{0.3\textwidth}
\input{full_frame.tex}
\end{scaletikzpicturetowidth}

\begin{scaletikzpicturetowidth}{0.4\textwidth}
\begin{tikzpicture}[xscale=.95,transform shape]
\draw[black,very thick] (-6.3,-2.1) rectangle (-2.1,2.1);
\draw[black,very thick] (-2.1,-2.1) rectangle (2.1,2.1);
\draw [step=.3, thick, draw= black, fill=black!10!green!30] (-6.3,-2.1) grid  (-2.1,1.6) rectangle (-6.3,-2.1);
\draw [step=.3, thick, draw= black, fill=black!10!blue!30] (-6.3,1.5) grid  (-2.1,2.1) rectangle (-6.3,1.5);
\draw [step=.3, thick, draw= black, fill=black!10!red!30] (-2.1,-2.1) grid  (2.1,0.4) rectangle (-2.1,-2.1);
\draw [step=.3, thick, draw= black, fill=black!10!yellow!30] (-2.1,0.3) grid  (2.1,1.2) rectangle (-2.1,0.3);
\draw [step=.3, thick, draw= black, fill=black!10!orange!30] (-2.1,1.2) grid  (2.1,2.1) rectangle (-2.1,1.2);

\draw [step=.3, thick, draw= black, fill=black!20!white!20] (-6.3,-2.1) grid  (-5.7,2.1) rectangle (-6.3,-2.1);
\draw [step=.3, thick, draw= black, fill=black!20!white!20] (-2.1,-2.1) grid  (-1.5,2.1) rectangle (-2.1,-2.1);

\node[align=center, scale =0.9] at (0,-2.5) {Slot 2};
\node[align=center, scale =0.9] at (-4.2,-2.5) {Slot 1};
\node[rotate=90, scale =0.9] at (-6.6,0) {Physical resource blocks (PRBs)};
\end{tikzpicture}
\end{scaletikzpicturetowidth}
\caption{5G time-sliced frame structure (top), consisting of 10 equal sized subframes of length 1 ms. A single subframe of each frame (dotted) is reserved for low-latency (LL) control system traffic. Within this subframe (bottom), users are scheduled into time-divided slots and frequency divided subcarriers---the number of each determined by a given \emph{numerology}. In the presented example, transmissions of $m=5$ plants (green, blue, red, yellow, orange) across $s=2$ slots and $b=14$ PRBs. Each slot contains 14 OFDM symbols, with the first two reserved for control overhead (gray).}
\label{fig_grid2}
\end{figure}

The 5G scheduling framework provides a flexible two dimensional grid-like architecture for scheduling transmissions. Such transmissions are slotted into a continuous series of frames---each of which consists of 10 equally sized subframes of length 1 ms. Each subframe moreover consists of 14 OFDM symbols, two of which we assume are reserved for overhead---see, e.g., \cite{mogensen20135g,lahetkangas2014achieving}. In modern IoT systems, standard broadband wireless traffic is mixed with transmissions associated with control systems fast dynamics or time critical features. This latter traffic has strict latency and reliability requirements that make it incompatible with standard broadband scheduling methods. Various features of 5G, such as network slicing \cite{popovski20185g} and puncturing \cite{anand2018joint}, have been proposed to meet service requirements of low latency traffic, requiring varying degrees of complexity. Given the periodic nature of low-latency control system transmissions, we propose a simple separation of broadband and low-latency traffic by subframe---i.e. a time-domain slicing of low latency and broadband traffic. E.g., with a 10 ms sampling rate, we may reserve one or two subframes out of every frame for the scheduling of low-latency traffic while leaving the network free for broadband traffic during the remaining time. This free time may also be used for both control signal overhead in scheduling as well as computational effort in determining optimal schedules. 

For the purposes of this paper and without loss of generality, we thus restrict our attention to a single subframe for the scheduling of low-latency traffic in every 10 millisecond long frame. Such an architecture ensures a 1 millisecond latency bound on the uplink of state information to the controller---a necessary requirement for many industrial control systems \cite{varghese2014wireless}. Within each subframe, the specific grid size is determined by a given \emph{numerology}. A specified numerology determines both the number $s$ of time-divided \emph{slots}---each of length $t_s := 1/s$ milliseconds---and $b$ physical resource blocks (PRBs)---each of bandwidth $f^{\prime}_b := 12f_b$, where $f_b$ is the so-called subcarrier spacing (SCS) of the given numerology. A higher index numerology will have more, shorter length slots and fewer, larger sized PRBs. We refer the reader to \cite{agiwal2016next,parkvall2017nr} for more details on the 5G frame structure. 

In addition to an assignment of slots and PRBs, each transmission is given a modulation and coding scheme (MCS). For user $i$, its MCS $\mu_i$ is selected from a discrete set of values $\ccalM:= \{0, 1, \hdots, \eta\}$. A higher MCS value indicates a higher modulation order and coding rate, which will generally increase the speed of transmission, or data rate, but with higher probability of packet error. To quantify these effects, we further consider a fading channel state, or effective SNR, vector for user $i$ at cycle $k$ as $\bbh_{i,k} \in \reals^b_+$, where $\bbh_{i,k}(j)$ is the fading channel gain in PRB $j$. We assume the channel coherence time is such that this state varies across control cycles but not within a single cycle. This is to say that $\bbh_{i,k}$ is constant within a single 5G frame. Given an MCS $\mu$ and channel state $\bbh$, we first define a function  $q(\mu, \bbh)$ which returns the probability of packet loss, otherwise called packet error rate (PER). Given the switched control dynamics in \eqref{eq_control_switch}, we equivalently say that $\gamma_{i,k} =0$ with probability $q(\mu_{i,k}, \bbh_{i,k})$. Likewise, we define $\tau(\mu,B)$ to be a function that, given an MCS $\mu$ and bandwidth $B$, returns the maximum time taken for a single transmission attempt. Both of these functions play a critical role in determining scheduling decisions in time-sensitive wireless control system settings. We are, in particular, interested in exploring the trade-off between PER and transmission time with different slot, PRB, and MCS selections. Generally speaking, the functions $q(\mu, \bbh)$ and $\tau(\mu,B)$ relate to $\mu$ by 
\begin{equation}\label{eq_mcs_r}
\mu' > \mu \implies q(\mu, \bbh) \leq q(\mu', \bbh), \quad \tau(\mu',B) \leq \tau(\mu,B).
\end{equation}
%

%Because a TD cannot finish until all transmissions within the TD have been completed, the total transmission time of a single TD $s$ is the maximum transmission time taken by all devices within that time slot. We define the transmission time of TD slot $s$ as
%%
%\begin{equation}\label{eq_time_slot}
%\hat{\tau}(\bbSigma, \bbmu, \bbalpha, s) := \max_{i: \alpha_i = s} \tau(\mu_i, \bbsigma_i) + \tau_0(\bbalpha,s),
%\end{equation}
%%
%where $\tau_0: \mathbb{Z}_{++}^m \times \mathbb{Z}_{++} \rightarrow \reals_+$ is a function that specifies the communication overhead of TD $s$. This overhead may consist of, e.g., the time required to send TFs to scheduled users, as seen in Figure \ref{fig_multiplex}.

%%%%%%%%%%%%%%%%%%%%%%%%%%%%%%%%%%%%
%%%%%%%%%%%%%%%%%%%%%%%%%%%%%%%%%%%%
%%%%  S  E  C  T  I  O  N     %%%%%%%%%%%%%%%%%%
%%%%%%%%%%%%%%%%%%%%%%%%%%%%%%%%%%%%
%%%%%%%%%%%%%%%%%%%%%%%%%%%%%%%%%%%%
\section{Optimal Scheduling}\label{sec_scheduling}

In each frame/control cycle, we are interested in scheduling the low-latency transmissions of plant states to the BS such as to maximize the performance of the system utilizing the time-sliced 5G framework in Fig. \ref{fig_grid2}---that is, scheduling low-latency control system uplinks in a single subframe of length 1 ms. The varying nature of both channel and plant states necessitates that the \emph{optimal} scheduling decision be adaptive to current conditions of the system. For notation convenience, we define the collection of all plant states $\bbX_k := [\bbx_{1,k}; \hdots; \bbx_{m,k}] \in \reals^{m \times p}$ and channels $\bbH_k := [\bbh_{1,k}; \hdots; \bbh_{m,k}] \in \reals^{m \times b}_{++}$ at cycle $k$.

To specify a scheduling decision, recall that a subframe can be divided as a $s \times b$ grid, with $s$ time slots and $b$ frequency resource blocks. Each plant may be scheduled in any set of adjacent grid points of the sub-frame; we denote this allocation for plant $i$ at cycle $k$ by a binary matrix $\bbA_{i,k} = \{0,1\}^{s \times b}$, where $\bbA_{i,k} = \bb0$ indicates no transmission. We further specify the set of grids with adjacent allocations as $\ccalA$ and denote by $l_{i,k}$ and $n_i$ the number of adjacent allocations in the horizontal (time-wise) and vertical (frequency-wise) directions, respectively. The lengths can equivalently be seen respectively as the total time and bandwidth of the scheduling allocation. To avoid collision, we impose that $\sum_i \bbA_{i,k} \leq \mathbf{1}\mathbf{1}^T$--at most one user may be assigned to any grid point. 

Furthermore, each transmission is given an MCS allocation $\mu_{i,k}$ that ultimately determines both its speed and probability of packet loss. Recalling the effect and MCS $\mu$ has on these metrics in \eqref{eq_mcs_r} and given the fixed size of a single slot---$t_s$---and fixed size of PRBs---$f^{\prime}_b$---it reasons to select the smallest $\mu_{i,k}$ such that the transmission time is completed by the end of its assigned slots; this ensures the smallest possible packet error rate. The MCS $\mu_i$ is then fully determined by its time- and frequency-wise allocation lengths $l_{i,k}$ and $n_i$ as
\begin{equation}\label{eq_mcs}
\mu_{i,k} := \mu(\bbA_{i,k}) = \min \{\mu \in \ccalM \mid \tau(\mu, n_i f^{\prime}_b) \leq l_{i,k} t_s \}.
\end{equation}

The optimal schedule is one that minimizes some cost or performance metric. Standard schedulers are ultimately interested in some measure of throughput. A natural cost to promote throughput is the expected number of successful transmissions per cycle. Given a set of scheduling assignments $\mathbb{A} = \{\bbA_i\}_{i=1}^m$, we may write this cost as 
\begin{align}\label{eq_simple_util}
\omega(\mathbb{A}, \bbH) := \E_{\bbH} \sum_i  \mathbbm{1}[\gamma_i = 0] = \sum_{i=1}^m q (\mu(\bbA_i),\bbh_i).
\end{align}
The optimal scheduling at cycle $k$ in the given 5G framework can be written as the solution to the problem
\begin{align}\label{eq_simple_problem}
\{\bbA_{1,k},\hdots,\bbA_{m,k}\} := \mathbb{A}^*(\bbH_k) =&\!\! \argmin_{\bbA_i,\hdots,\bbA_m \in \ccalA} \omega(\mathbb{A},\bbH_k), \\
&\st \quad \sum_{i=1}^m \bbA_{i} \leq \mathbf{1}\mathbf{1}^T. \nonumber
\end{align}
In \eqref{eq_simple_problem}, we minimize the cost in \eqref{eq_simple_util} under current channel conditions $\bbH_k$, subject to a constraint that restricts a single user to be scheduled in each slot or PRB. The problem in \eqref{eq_simple_problem} is generally hard to solve, due to the discrete and combinatorial sized search space. However, if we restrict the scheduling assignments to a \emph{single} grid point---i.e. each user is given a single slot and a fixed number $\rho$ PRBs---then \eqref{eq_simple_problem} reduces to a standard linear-sum assignment problem with $n := s \lfloor b/\rho \rfloor $ distinct and mutually exclusive assignments, each incurring a cost of $q (\mu(\bbA_i),\bbh_i)$ In this case, the solution $\mathbb{A}^*(\bbH_k)$ can be found using the Hungarian method \cite{kuhn1955hungarian}, or a low complexity greedy approximation \cite{romeijn2000class,kang2011task,thekumparampil2015combinatorial}. Thus, at every cycle, given the fading channel states $\bbH_k$, the optimal schedule is found by determining the slot/PRB assignments $\{\bbA_{1,k},\hdots,\bbA_{m,k}\}$ via \eqref{eq_simple_problem}. The MCS selections then follow from \eqref{eq_mcs}.

The optimal schedules found via \eqref{eq_simple_problem} may not, in practice, be enough to obtain strong performance over the tight latency constraints required. By limiting to a  single subframe we are placing a potentially severe restriction on the number of plants that can be serviced in each cycle. As modern IoT system grow in size, the more traditional utility measure of \eqref{eq_simple_util} is not the most appropriate to adequately utilize the limited radio resources available. Observe that the optimal design problem \eqref{eq_simple_problem} seeks to maximize the total number of successful transmission across the systems, but agnostic to the control plant states and dynamics. Thus, it does not capture any sense of priority or urgency with regard to certain plants over others.  Indeed, an important insight to consider is the fact that we are ultimately interested in the performance of the control systems; the performance of the underlying communications network is only of interest in as far as it effects the plant dynamics in \eqref{eq_control_switch}. We proceed in the next section to derive a more accurate utility and optimal scheduling design problem for 5G architectures.

\subsection{Control optimal scheduling}\label{sec_scheduling_control}

Incorporating the switch system plant dynamics in \eqref{eq_control_switch} and standard measure of control performance, we may construct a control-aware variation of the optimal scheduling problem in \eqref{eq_simple_problem}. Consider the standard Lyapunov quadratic cost function $L(\bbx) := \bbx^T \bbP \bbx$ for a given positive definite matrix $\bbP \in \reals^{p \times p}$. The quadratic cost $L(\bbx)$ measures the current cost of a control system in state $\bbx$---e.g. if $\bbx$ represents some distance from desired operating point, we may set $\bbP = \bbI$ and have $L(\bbx)$ reflect the Euclidean norm. We are then, after all, interested in designing 5G scheduling policies that achieve good performance across all systems, which can be measured with as the sum $\sum_i L(\bbx_{i,k}$ at time $k$.

From here, we show how a scheduling decision impacts the control performance of each plant. Wireless communications are subject to random packet drops given the channel state, which ultimately effect the dynamic evolution of the plants. Recall the packet error rate experience by system $i$ given a scheduling assignment $\bbA_i$ and channel state $\bbh_i$ given by $q(\mu(\bbA_i),\bbh_i)$. Using the two cases of dynamics given in \eqref{eq_control_switch}, we may consider the \emph{expected} quadratic cost under such a packet error rate given the known or estimated state $\hbx^{(l_{i,k})}_{i,k}$ as
\begin{align}\label{eq_expected_cost}
\E_{\bbq, \bbw} &\left\{ L(\bbx_{i,k+1}) \mid \bbA_i, \bbh_i, \hbx^{(l_{i,k})}_{i,k} \right\} =  \\
& \E_{\bbw} \Big\{ 
(1-q(\mu(\bbA_i),\bbh_i))  L\left( (\bbA_i + \bbB_i \bbK_i) \bbx_{i,k} + \bbw_k \right)   \nonumber  \\
&  +
q(\mu(\bbA_i),\bbh_i) L\left( \bbA_i \bbx_{i,k} + \bbB_i\bbK_i\hbx^{(l_{i,k})}_{i,k} + \bbw_k \right) \Big\}. \nonumber
\end{align}
The expectation in the left hand side of \eqref{eq_expected_cost} is with respect to the random packet loss and system noise $\bbw$, while the expectation in the right hand side is only with respect to $\bbw$. Given states $\bbh_i$ and $\hbx^{(l_{i,k})}_{i,k}$ and scheduling assignment $\bbA_i$, the expected cost at the next step is a convex combination of the two possible evolutions of $\bbx_i$ under such a scheduling. We may further reduce the cost in \eqref{eq_expected_cost} taking the expectation over $\bbw_k$ and reducing as
\begin{align}\label{eq_expected_cost_r}
\E_{\bbq, \bbw} &\left\{ L(\bbx_{i,k+1}) \mid \bbA_i, \bbh_i, \hbx^{(l_{i,k})}_{i,k} \right\} =  \\
& 
\left[  \| \bbA_i^c \hbx\|^2_{\bbP^{\frac{1}{2}}}+ \Tr(\bbP \bbW) + \sum_{j=1}^{l}\omega_i^{j} \right] +  \nonumber  \\
&  
q(\mu(\bbA_i),\bbh_i) \sum_{j=0}^{l_{i,k}-1}\left[ \Tr( \bbA_{i}^{cT} (\bbA_i^T\bbP^{\frac{1}{j}}\bbA_i)^j \bbA_i^c \bbW) - \omega_i^{j+1}\right], \nonumber
\end{align}
where $\bbA_i^c := \bbA_i + \bbB_i\bbK_i$ and $\omega_i^j := \Tr[(\bbA_i^T\bbP^{1/j}\bbA_i)^{j} \bbW]$ are defined for notational convenience. We point out that algebraic details of the reduction from \eqref{eq_expected_cost} to \eqref{eq_expected_cost_r} can be found in the proof of \cite[Proposition 1]{eisen2019control}.

The complete reduced form of the expected control cost in \eqref{eq_expected_cost_r} reveals how the control-aware scheduling problem may be formulated. Firstly, observe that the effect of a scheduling assignment $\bbA_i$ through the packet error rate $q(\mu(\bbA_i),\bbh_i)$ seen in the rightmost term in \eqref{eq_expected_cost_r} is not impacted by the estimated state value $\hbx^{(l_{i,k})}_{i,k}$. This implies that a scheduling decision does not require any estimation of the state, and only needs to know the counter $l_{i,k}$ of the last successful transmission.  Thus, by removing the first term in the right hand side of \eqref{eq_expected_cost_r} that does not depend upon $\hbx^{(l_{i,k})}_{i,k}$, we may consider the a cost of a complete scheduling $\mathbb{A}$ that sums the expected costs across all systems given channel conditions $\bbH$ and transmission counters $\bbl_k := [l_{1,k}; \hdots; l_{m,k}]$ as
\begin{align}\label{eq_control_util}
\omega^{\prime}(\mathbb{A}, \bbH, \bbl_k) := \sum_{i=1}^m c_i(l_{i,k}) q (\mu(\bbA_i),\bbh_i),
\end{align}
where we have defined the control system weighting term for system $i$ as 
\begin{align}
c_i(l_{i,k}) :=  \sum_{j=0}^{l_{i,k}-1}\left[ \Tr( \bbA_{i}^{cT} (\bbA_i^T\bbP^{\frac{1}{j}}\bbA_i)^j \bbA_i^c \bbW) - \omega_i^{j+1}\right].
\end{align}
The control optimal scheduling at cycle $k$ in the given 5G framework can then be written as the solution to the problem
\begin{align}\label{eq_control_problem}
\{\bbA_{1,k},\hdots,\bbA_{m,k}\} :=&\argmin_{\bbA_i,\hdots,\bbA_m \in \ccalA}\omega^{\prime}(\mathbb{A},\bbH_k, \bbl_k), \\
&\st \quad \sum_{i=1}^m \bbA_{i} \leq \mathbf{1}\mathbf{1}^T. \nonumber
\end{align}

In \eqref{eq_control_problem} the scheduling allocation is determined so as to minimize the summation of the expected quadratic costs in \eqref{eq_expected_cost} for all plants $i=1,\hdots,m$. This reduces to a linear scaling of the packet delivery rate $q (\mu(\bbA_i),\bbh_i)$ by a weighting term $c_i(l_{i,k}) <0$ for each system $i$, that takes into accounts the dynamics of plant $i$ and its last transmission counter $l_{i,k}$. %We stress that, contrary to the utility \emph{maximization} performed in \eqref{eq_simple_problem}, we are \emph{minimizing} a cost in \eqref{eq_control_problem}.  
As in the previous case, when restricting allocation to a single slot and $\rho$ PRBs for each user, \eqref{eq_control_problem} becomes a standard linear cost assignment problem whose solution can be found using the Hungarian method \cite{kuhn1955hungarian} or a low complexity greedy approximation \cite{romeijn2000class,kang2011task,thekumparampil2015combinatorial}. Note that, when $m > n := s \lfloor b/\rho \rfloor$, not all users can be scheduled in each frame.

The full scheduling algorithm can be summarized as follows. At each cycle $k$, the plant states are sampled locally at each device to be transmitted. In the current 5G transmission frame, we reserve a single subframes of length 1 ms to schedule for a total of $n = s \lfloor b/\rho \rfloor$ scheduling assignment blocks. The scheduling decision is then made with the routine:
\begin{enumerate}
\item Measure channel states $\bbH_k = [\bbh_{1,k}; \hdots; \bbh_{m,k}] \in \reals^{m \times b}_{++}$.
\item Determine scheduling alloc. $\{\bbA_{1,k},\hdots,\bbA_{m,k}\}$ via \eqref{eq_control_problem}.
\item Set MCS values $\mu_{i,k}$ via \eqref{eq_mcs} for all $i=1,\hdots,m$.
\item Device $i$ transmits with packet success $\gamma_{i,k}=1$ w.p. $q(\mu_{i,k}, \bbh_i)$ for all $i=1,\hdots,m$.
\item BS applies corresponding control and plant $i$ evolves via \eqref{eq_control_switch} for all $i=1,\hdots,m$.
\item Counter $l_{i,k}$ updates via \eqref{eq_time_switch} for all $i=1,\hdots,m$.
\end{enumerate}

%%%%%%%%%%%%%%%%%%%%%%%%%%%%%%%%%%%%%%%%%%%%%%%%%%%%%%%%%%%%%%%%%
%%%%%   A   L   G   O   R   I   T   H   M   %%%%%%%%%%%%%%%%%%%%%
%%%%%%%%%%%%%%%%%%%%%%%%%%%%%%%%%%%%%%%%%%%%%%%%%%%%%%%%%%%%%%%%%
%{\linespread{1.3}
%\begin{algorithm}[t] \begin{algorithmic}[1]
%\STATE \textbf{Parameters:} Lyapunov decrease rate $\rho$
%\STATE \textbf{Input:} Channel conditions $\bbh_{i,k}$ and estimated states $\hbx^{(l_{i,k})}_{i,k}$ for all $i$
%\STATE Compute target PDR $\tdq_i(\hbx^{(l_{i,k})}_{i,k})$ for each device $i$ [cf. \eqref{eq_pdr_constraint}].
%\STATE Determine selection probabilities $\nu_{i,k}$ for each device [cf. \eqref{eq_prob_c}].
%\STATE Select devices $\ccalI_k$ with probs. $\{\nu_{1,k},\hdots,\nu_{m,k}\}$
%\STATE Determine set of FDs/TDs $\ccalS'_k$ [cf. \eqref{eq_ru_sets}].
%\STATE Determine max. DR for each device/FD assignment [cf. \eqref{eq_mcs_select}].
%\STATE Schedule selected devices via assignment method \cite{kuhn1955hungarian}.
%\STATE \textbf{Return:} Scheduling variables $\{\bbsigma_{i}, \mu_i, \alpha_i\}_{i=1}^m$ 
%\end{algorithmic}
%\caption{Control-aware scheduling for low-latency at cycle $k$}\label{alg_calls} \end{algorithm}}
%%%%%%%%%%%%%%%%%%%%%%%%%%%%%%%%%%%%%%%%%%%%%%%%%%%%%%%%%%%%%%%%%

  %%%%%%%%%%%%%%%%%%%%%%%%%%%%%%%%
%%%%%%%%%% F I G U R E %%%%%%%%%%%%%%%%%
%%%%%%%%%%%%%%%%%%%%%%%%%%%%%%%%

\begin{figure}[t]
\centering
\includegraphics[height=.3\textheight, width=.45\textwidth]{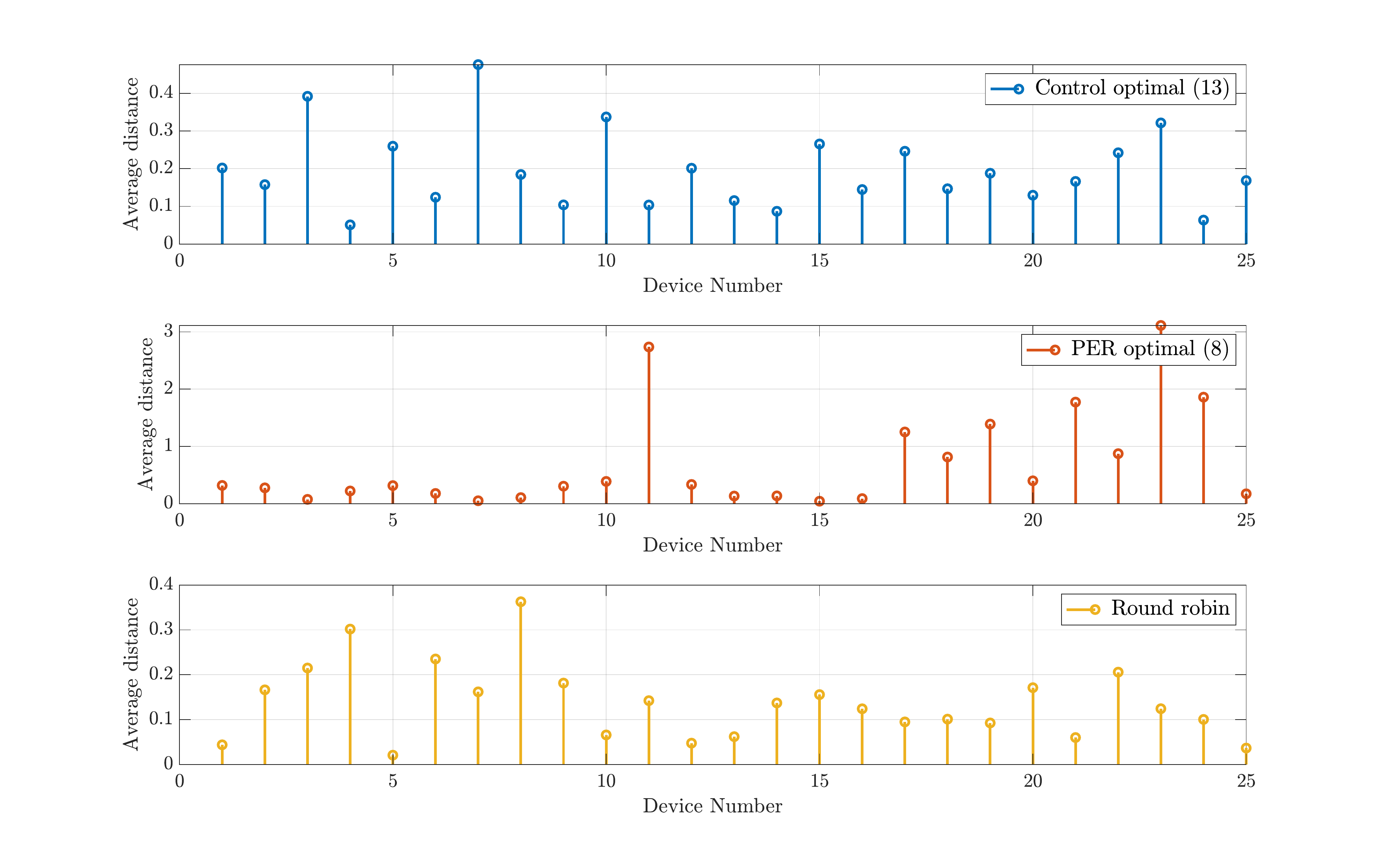}
\caption{Average distances from center vertical for $m=25$ pendulums using different scheduling methods for a single trial. The control-optimal scheduling and round robin keep the pendulums while the PER optimal scheduling does not. }\label{fig_sim_results2}
\end{figure}

\begin{figure}[t]
\centering
\includegraphics[height=.2\textheight, width=.4\textwidth]{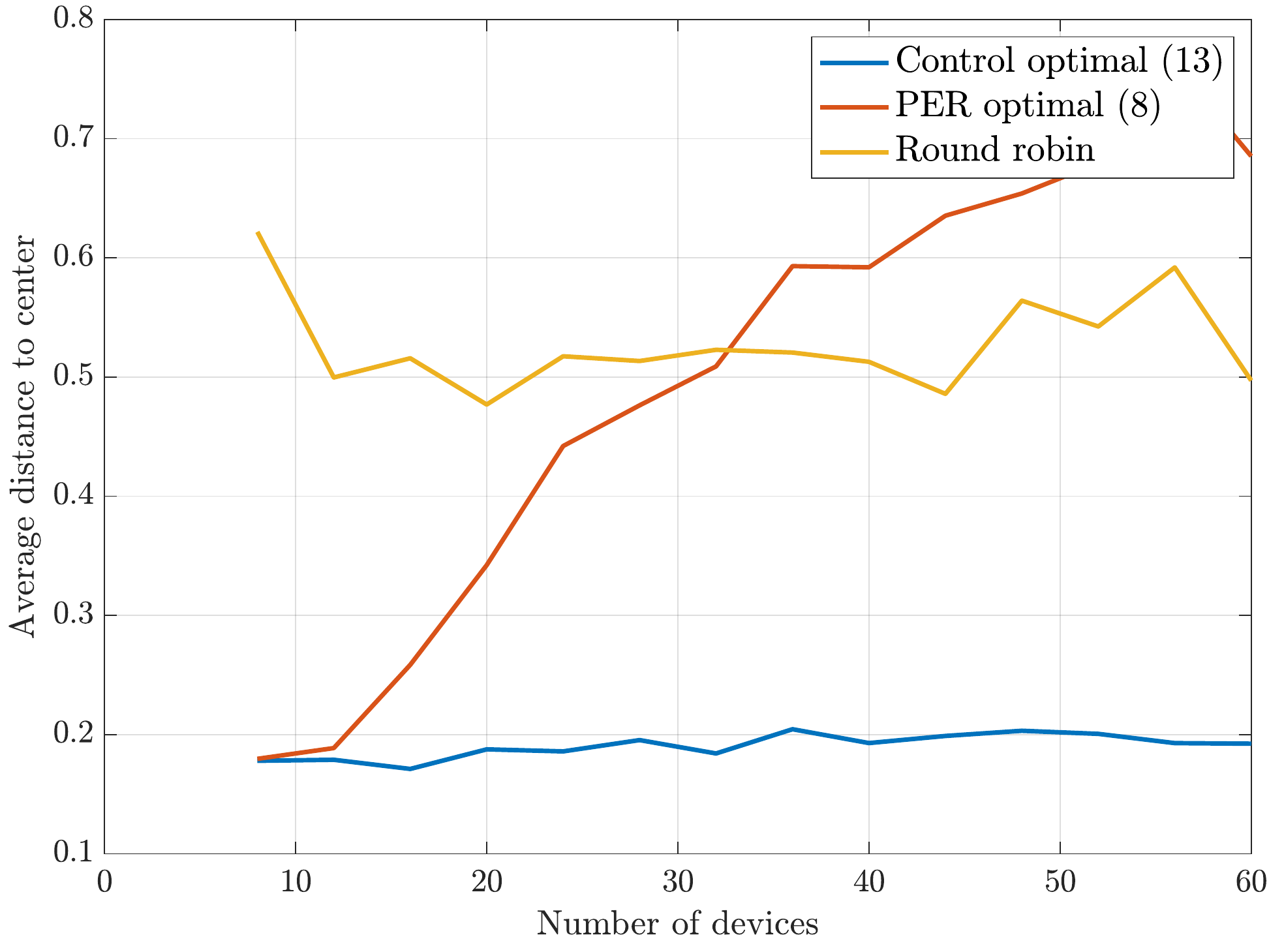}
\caption{The average distances from center vertical for increasing number of pendulums over 10 independent trials. The control-optimal scheduling maintains a small distance from center up to 60 total devices, while the PER optimal scheduling gets worse as system size grows.}\label{fig_sim_results}
\end{figure}

%The complete control-aware scheduling procedure for low-latency settings is present in Algorithm \ref{alg_calls}. At each cycle $k$, the BS uses current channel states $\bbh_{i,k}$ (obtained via pilot signals) and the current estimated control states $\hbx^{(l_{i,k})}_{i,k}$  (obtained via \eqref{eq_state_est} for each device $i$) to compute control-aware target PDRs  $\tdq_i(\hbx^{(l_{i,k})}_{i,k})$ for each device via \eqref{eq_pdr_constraint} in Step 3. In Step 4, the target PDRs are used to establish selection probabilities $\nu_{i,k}$ for each agent with \eqref{eq_prob_c}. After randomly selecting devices $\ccalI_k$ in Step 5, an appropriate set of FDs and TDs $\ccalS'_k$ are selected in Step 6. In Step 7, the associated DR values are determined for each possible assignment of device to FDs via \eqref{eq_mcs_select}. Finally, in Step 8 the assignment is performed using, e.g., the Hungarian method or other user-designed heuristic assignment method. The resulting assignment determines the scheduling parameters $\{\bbsigma_i, \mu_i, \alpha_i\}$ for all devices $i$ in the current cycle. 

 %%%%%%%%%%%%%%%%%%%%%%%%%%%%%%%%%%%%%%%%%%%%%%%%%%%%%%%%%%%%%%%%%%%%
%%%   S   E   C   T   I   O   N   %%%%%%%%%%%%%%%%%%%%%%%%%%%%%%%%%%
%%%%%%%%%%%%%%%%%%%%%%%%%%%%%%%%%%%%%%%%%%%%%%%%%%%%%%%%%%%%%%%%%%%%
 
 \section{Simulation Results}\label{sec_numerical_results}

To evaluate the performance of the control-optimal scheduling procedure outlined in the previous section, we perform numerical simulations on the low latency control problem of maintaining a series of inverted pendulums on a horizontal cart over a wireless channel. The highly unstable dynamics of the inverted pendulum make it a representative example of control system that requires fast control cycles, and subsequently low-latency communications when being controlled over a wireless medium. Consider a series of $m$ identical inverted pendulums---where each pendulum is attached at one end to a cart that can move along a single, horizontal axis---using the modeling of the inverted pendulum. The state is $p=4$ dimensional vector that maintains the position and velocity of the cart along the horizontal axis, and the angular position and velocity of the pendulum, i.e. $\bbx_{i,k} := [x_{i,k}, \dot{x}_{i,k}, \theta_{i,k}, \dot{\theta}_{i,k}]$. The system input $u_{i,k}$ reflects a horizontal force placed upon $i$th pendulum. By applying a zeroth order hold on the continuous dynamics with a state sampling rate of $10$ milliseconds and linearizing, we obtained the following discrete linear dynamic matrices of the pendulum system $\bbA_i$ and $\bbB_i$.
%%
%\begin{align}\label{eq_control_orig1}
%\bbA_i =
%\begin{bmatrix}
%1 & 0 & 0 & 0 \\
%0 & 2.055 & -0.722 & 4.828 \\
%0 & 0.023 & 0.91 & 0.037 \\
%0 & 0.677 & -0.453 & 2.055
%\end{bmatrix},
%\bbB_i =
%\begin{bmatrix}
%0.034 \\ 0.168 \\ 0.019 \\ 0.105
%\end{bmatrix}.
%\end{align}
%%
Because the state $\bbx_{i,k}$ measures the angle of the $i$th pendulum at time $k$, the goal is to keep this close to zero, signifying that the pendulum remains upright.  We perform the scheduling using the proposed control-optimal formulation in \eqref{eq_control_problem}, the packet error rate (PER)-optimal scheduling formulation \eqref{eq_simple_problem}, and a baseline round robin scheduling procedure. Each simulation is run for a total of $1000$ seconds using an LQR controller. Our simulation environment consists of a 20MHz channel with devices dropped in a 50 m radius from the BS. The numerology includes 2 slots of length 500 $\mu$s and a subcarrier spacing of 30 kHz, with each user allocated $n=12$ PRBs.

In Figure \ref{fig_sim_results}, we show the average distance of each of $m=25$ pendulums in a trial simulation using each of the three scheduling methods. Both the control-optimal scheduling and the round robin keep the pendulums upright, while the PER-optimal scheduling is not able to keep all the pendulums. For a more comprehensive view of the performance, we perform simulations with an increasing number of devices with multiple trials. In Figure \ref{fig_sim_results} we show the average distance across 10 independent trials to the center across all the pendulums for increasing number of total devices sharing the 5G network. As can be seen, as the number of devices increase, the PER optimal schedule has a harder time keep the pendulums close to the center, while the control-optimal scheduling maintains a steady distance from the center even as the number of devices increase. The round robin scheduler is consistent across network size but cannot keep the pendulums close to center. This demonstrates that, in the control optimal scheduling, the consideration of the pendulum dynamics and the control state helps make stronger scheduling decisions in keeping the systems in good condition.

 %%%%%%%%%%%%%%%%%%%%%%%%%%%%%%%%%%%%%%%%%%%%%%%%%%%%%%%%%%%%%%%%%%%%
%%%   S   E   C   T   I   O   N   %%%%%%%%%%%%%%%%%%%%%%%%%%%%%%%%%%
%%%%%%%%%%%%%%%%%%%%%%%%%%%%%%%%%%%%%%%%%%%%%%%%%%%%%%%%%%%%%%%%%%%%
 
 \section{Conclusion}
In this paper we develop a control-optimal approach towards scheduling for low-latency, or time sensitive, wireless control systems in 5G networks. Because many control systems in industrial settings require very low latency transmission to operate effectively, there is an intrinsic challenge in trading off the data rates necessary to achieve low latency with the packet error rates necessary for high reliability. We propose a simple time-slicing architecture that reserves a single or couple of subframes in each 5G frame for the scheduling of low-latency traffic used for the control systems. We further demonstrate that by allocating radio resources relative to underlying control system dynamics and states, we can more effectively utilize the limited scheduling resources available in the available subframes in low-latency 5G systems. Numerical results demonstrate stronger performance relative to control-agnostic scheduling in terms of the number of system that can be supported by the communication network.

\urlstyle{same}
\bibliographystyle{IEEEbib}
\bibliography{wireless_ll_control,scheduling_control}

\end{document}

%% file: my_sections.tex
\usepackage{needspace}

%% file: tikz_wncs4.tex
\pgfdeclarelayer{bg0}    % declare background layer
\pgfdeclarelayer{bg1}    % declare background layer
\pgfsetlayers{bg0,bg1,main}  % set the order of the layers (main is the standard layer)

\tikzstyle{block} = [draw,rectangle,thick,
text height=0.2cm, text width=0.7cm, 
fill=blue!30, outer sep=0pt, inner sep=0pt]
\tikzstyle{dots} = [font = \large, minimum width=2pt]
\tikzstyle{dash_block} = [draw,rectangle,dashed,minimum height=1cm,minimum width=1cm]
\tikzstyle{smallblock} = [draw,rectangle,minimum height=0.5cm,minimum width=0.5cm,fill= green!30, font =  \scriptsize]
\tikzstyle{smallcircle} = [draw,ellipse,minimum height=0.1cm,minimum width=0.3cm,fill= yellow!40, font =  \scriptsize ]
\tikzstyle{connector} = [->]
\tikzstyle{dash_connector} = [->,thick,decorate,decoration={snake, amplitude =1pt, segment length=8pt}, magenta]
\tikzstyle{branch} = [circle,inner sep=0pt,minimum size=1mm,fill=black,draw=black]

\tikzstyle{vecArrow} = [thick, decoration={markings,mark=at position
   1 with {\arrow[semithick]{open triangle 60}}},
   double distance=1.4pt, shorten >= 5.5pt,
   preaction = {decorate},
   postaction = {draw,line width=1.4pt, white,shorten >= 4.5pt}]

\begin{tikzpicture}[scale=1, blocka/.style ={rectangle,text width=0.9cm,text height=0.6cm, outer sep=0pt}]
 \small

    % node placement with matrix library: 5x4 array
    \matrix(M)[ampersand replacement=\&, row sep=2.0cm, column sep=10pt] {
    
    %\&
    \node[smallblock, align=center] (CS1) {Control \\ System {1}};\&\&
    \node[smallblock, align=center] (CS2) {Control \\ System {2}};\&\&\&
%    \&
    \node(d1) {$\cdots$};\&
%    \&
    \node[smallblock, align=center] (CSm) {Control \\ System \textit{m}};\&
    \\
    \node[blocka] (R1) {};\&\&
    \node[blocka] (R2) {};\&\&\&
%    \node[smallcircle] (R2) {R2};\&
    \node[blocka] (d3) {};\&
    \node[blocka] (Rm) {};\&
    \\
    };

    \node[block] (outer) [fit=(R1.north west) (d3) (Rm.south east)] {};
    
    \node[align=center, scale =0.9] at (outer.center) {Access Point/ \\Controller};
    
    \draw [->, thick, red] (CS1) -- node[left]{} (R1);
    \draw [->, thick, red] (CS2) -- node[left]{} (R2);
%    \draw [->, thick, magenta] (T2) -- node[left]{ \scriptsize $h_2$} (R2);
    \draw [->, thick, red] (CSm) -- node[left]{} (Rm);

		\begin{pgfonlayer}{bg0}    % select the background layer
		\draw [->, dashed, black] (R1) |- ($(R1) + (+35pt,-20pt)$) node(down_right){} 
		-- ($(CS1) + (+35pt,+20pt)$) node(up_right){} -| (CS1);
		\end{pgfonlayer}

		\begin{pgfonlayer}{bg0}    % select the background layer
		\draw [->, dashed, black] (R2) |- ($(R2) + (+35pt,-20pt)$) node(down_right){} 
		-- ($(CS2) + (+35pt,+20pt)$) node(up_right){} -| (CS2);
		\end{pgfonlayer}

		\begin{pgfonlayer}{bg0}
		\draw [->, dashed, black] (Rm) |- ($(Rm) + (+35pt,-20pt)$) node(down_right){} 
		--($(CSm) + (+35pt,+20pt)$) node(up_right){} -| (CSm);
		\end{pgfonlayer}

		\begin{pgfonlayer}{bg1}
		%\begin{scope}[on background layer]
		\node(shared) [fill=red!10, fit={($(CS1.south) + (-15pt, -10pt)$) 
		($(CS2.south) + (-10pt, -10pt)$)
		($(CSm.south) + (+20pt, -10pt)$)
		($(R1.north) + (-15pt, +10pt)$)
		($(R2.north) + (-10pt, +10pt)$)
		($(Rm.north) + (+20pt, +10pt)$)
		}] {};
		%\end{scope}
		\end{pgfonlayer}
		
		\node[align=center, red!50](shared_medium) at (shared.center) {Shared \\ Wireless \\ Medium};

\coordinate (FIRST NE) at (current bounding box.north east);
   \coordinate (FIRST SW) at (current bounding box.south west);

	\pgfresetboundingbox
   \useasboundingbox ($(FIRST SW) + (+30pt,0)$) rectangle (FIRST NE);

\end{tikzpicture}

%% file: full_frame.tex
\begin{tikzpicture}[xscale=.95,transform shape]

\draw [-latex](-0.5,0) coordinate(dd)-- (0,0) coordinate (O1) -- (8,0)coordinate(ff) node[right]{$ms$};
\draw [dashed,thick] (O1) -- (0,1) coordinate(S3) -- (0,3) coordinate(S2) -- (0,5) coordinate(e3) -- ++(0,1)coordinate(ff2);

%\foreach \nn in{S3,S2,e3}{
%\draw [thick] (dd|-\nn) node[above]{\nn}-- (\nn-|ff);
%}

\foreach \xx in{1,2,...,10}{
\draw[dashed] (\xx*0.8,0) -- (\xx*0.8,0|- ff2);
}

\foreach \xx in{0,2,4,...,10}{
\draw[dashed] (\xx*0.8,0.2) -- (\xx*0.8,-0.2) node[below]{\xx };
}

\node[rotate=90, scale =0.9] at (-.5,3) {Frequency allocation};
\node[scale =0.9] at (5.5*0.8,3) {Broadband traffic};
\node[scale =0.9] at (2*0.8,5) {LL traffic};
\draw [-latex](2*0.8,4.8) coordinate(dd5)-- (0.4,4) coordinate (O16);

\begin{scope}[shift={(O1)}]
\node[ above right=0.0cm and 0cm of S2,right,draw, minimum width=0.8cm,minimum height=6cm,pattern=dots](n3a) {$$};
\end{scope}

\draw [-latex](0.4,3) coordinate(dd2)-- (0,-2) coordinate (O12);
\draw [-latex](0.4,3) coordinate(dd3)-- (8,-2) coordinate (O13);

\end{tikzpicture}